\documentclass[aps,prb,twocolumn,showpacs,amsmath,amssymb,superscriptaddress]{revtex4}
\usepackage{graphicx}
\usepackage{bm}

\def\be{\begin{equation}}       \def\ee{\end{equation}}
\def\bea{\begin{eqnarray}}      \def\eea{\end{eqnarray}}

\begin{document}

\title{D-Wave Checkerboard Order in Cuprates}
\author{Kangjun Seo}
\affiliation{Department of Physics, Purdue University, West
Lafayette, Indiana 47907}
\author{Han-Dong Chen}
\affiliation{Department of
Physics, University of Illinois at Urbana-Champaign, Urbana, IL
61801}
\author{Jiangping Hu}
\affiliation{Department of Physics, Purdue University, West
Lafayette, Indiana 47907}

\date{\today}

\begin{abstract}

We show that a $d$-wave ordering in particle-hole channel, dubbed as
$d$-wave checkerboard order, possesses important physics that can
sufficiently explain the STM results in cuprates. A weak $d$-wave
checkerboard order can effectively suppress the coherence peak in
the single-particle spectrum while leaving the spectrum along nodal
direction almost unaffected. Simultaneously,  it generates a Fermi
arc with little dispersion around nodal points at finite temperature
that is consistent with the results of ARPES experiments in the
pseudogap phase. We also show that there is a general complementary connection between the $d$-wave checkerboard order and the   pair density wave
order.  Suppressing superconductivity locally or globally through
phase fluctuation should induce both orders in underdoped cuprates
and explain the nodal-antinodal dichotomy observed in ARPES and STM
experiments.
\end{abstract}

\pacs{74.25.Jb, 74.25.Dw, 74.75.-h }

\maketitle

Recently, Scanning Tunneling Microscopy(STM) has revealed surprising
yet important electronic structures in  the high temperature
superconductors. The fourier transform scanning tunneling
spectroscopies (FT-STS) from STM  have captured two different
general features in both momentum and energy
spaces~\cite{HOFFMAN2002,HOWALD2003,HANAGURI2004,MCELROY2003,MCELROY2003A,VERSHININ2004,MCELROY2005A,MCELROY2005,
FANG2006}. One feature is dispersive peaks in
FT-STS~\cite{MCELROY2003, MCELROY2003A}, interpreted as interference
patterns caused by elastic scattering of quasiparticles from
impurities~\cite{WANG2003}. The other is non-dispersive peaks,  a
checkerboard modulation observed in various different materials and
circumstances.
 The checkerboard structure was first discovered locally in BSCCO near a vortex core.\cite{PAN2000,HOFFMAN2002}
  Then, it was found to be a characteristics of the large gap regions
  where the STM spectrum resembles that in the pseudogap phase.\cite{HOWALD2003,MCELROY2003A,MCELROY2005A}
  Later, the STM studies of Ca$_{2-x}$Na$_x$CuO$_2$Cl$_2$ revealed the presence of a global
  long range commensurate checkerboard order independent of doping.\cite{HANAGURI2004}
  Finally, in the pseudogap phase, a similar checkerboard pattern was also observed.\cite{VERSHININ2004}

   The origin of the checkerboard has become central to understanding
the nature of electronic states in cuprates. Various different
mechanisms have been considered to explain the observed
non-dispersive checkerboard modulation, including pair density
modulation~\cite{CHEN2002,CHEN2004,CHEN2004A,TESANOVIC2004,TESANOVIC2005},
current density modulation~\cite{BENA2004,GHOSAL2005}, spin
modulation~\cite{SACHDEV2004} , stripe charge
modulation~\cite{KIVELSON2003,ROBERTSON2006}, and impurity
scattering~\cite{PODOLSKY2003} and so on.  Among the proposed
mechanisms, the pair density wave (PDW) has been shown to capture
important characteristics of the checkerboard density modulation.
The mechanism of PDW derives from high pairing energy scale in
cuprates. It suggests that unlike the superconductivity of normal
BCS type superconductors that can be destroyed by breaking Cooper
pairs, the superconductivity in cuprates can be more easily weakened
or destroyed by phase fluctuations than by pair breaking. Based on
this argument,  pair density localization~\cite{CHEN2002} was first
proposed to explain the local checkerboard modulation in the
presence of impurity or vortex. Later, a global pair density
wave(PDW) was proposed to explain the checkerboard physics in the
pseudogap state~\cite{VERSHININ2004,CHEN2004}. It has also been
shown that the symmetry of the tunneling intensity can distinguish
the pair density modulation from the conventional density
modulation~\cite{CHEN2004}. While the pair density modulation
provides a good understanding of the experimental results, it does
not establish a direct link between superconducting and pseudogap
states, as is suggested by the presence of the non-dispersive
checkerboard density modulation in both states. Furthermore, the
theory has not explained three important features present in the STM
spectrum: {\bf 1.} the density of state at low energy in
superconducting state does not change whether or not a checkerboard
modulation takes place locally; {\bf 2.} the overall intensity of
the modulation is rather small; {\bf 3.} the small modulation has a
large effect on the STM spectrum around the superconducting gap.

 In this Letter, we show that an explanation based on $d$-wave checkerboard density (DWCB) order in particle-hole
channel can overcome the above limitations of the PDW theory. The
DWCB can be viewed as a natural extension of the d-density wave(DDW)
order proposed to explain pseudogap
physics\cite{NAYAK2000,CHAKRAVARTY2001}, and is only different from
the latter in terms of order wavevectors. We show that the DWCB
order must exist when the PDW order is present in the global $d$-wave
superconducting state. Moreover, we demonstrate that the DWCB, with
$\mathbf{\mathbf{Q}}=\{(\pi/2,0),(0,  \pi/2)\} $ and $f(\mathbf{k})=\cos (k_x)-\cos (k_y)$, can
explain the STM experimental results. We show that  the DWCB order
has little effect on the density of state at low energy in the
superconducting phase, but has a strong effect on the STM spectrum
around  the superconducting gap at high energy. This result natually
explains the puzzling dichotomy between the nodal and antinodal
regions observed in STM\cite{MCELROY2005A} and angle resolved
photoemission spectroscopy (APRES)~\cite{ZHOU2004}. The DWCB order
also preserves in FS-STM spectrum at the wavevector $\mathbf{Q}$ the same
symmetry as that observed in the experiments. Moreover, the DWCB
preserves the nodes in the single particle spectrum, and generates a
Fermi arc with little dispersion around the nodal points at high
temperature, which are consistent with the results from  ARPES. The
Fermi arc has been a signature of the pseudogap region, and has been
proposed to explain the checkerboard pattern observed in the
pseudogap state \cite{Chatterjee2006}. Thus, the DWCB provides a
physical origin of the Fermi arc. To our knowledge, this letter
presents the first concrete model describing this physics.

\begin{figure}
\includegraphics[width=7cm]{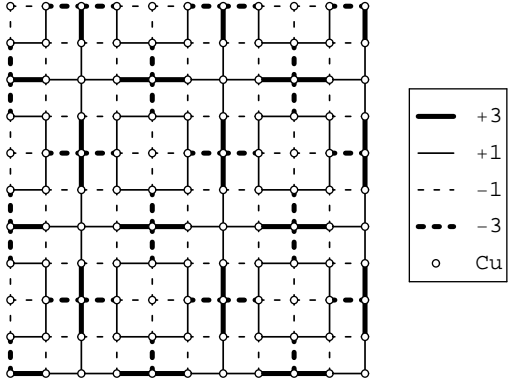}
\caption{\label{fig_1} The configuration of bond density of DWCB
order, or $W_\delta (r)$.
It is manifestly shown that the pattern has $4a \times 4a$
periodicity and $d_{x^2-y^2}$ symmetry.}
\end{figure}

{\it The connection between particle-particle (P-P) and
particle-hole (P-H) channel orders:} To illustrate the complementary connection, we make use of the DDW
order, since the only difference between the DWCB and DDW orders is
the order wavevectors. Let's consider a state with a DDW order,
$\langle \sum_\sigma c^+_{\mathbf{k}\sigma}c_{\mathbf{k}+\mathbf{Q}'\sigma}\rangle = i\Delta_{\text{ddw}}f(\mathbf{k})$, and a
$d$-wave superconducting order(DSC),
$\langle c_{\mathbf{k}\uparrow}c_{-\mathbf{k}\downarrow}\rangle =\Delta_{\text{dsc}}f(\mathbf{k})$, where
$\mathbf{Q}'=(\pi,\pi)$. Some simple calculation will show that in the above
mixed state of DDW and DSC, there naturally exists a PDW order with
a wavevector at $\mathbf{Q}'$, given by
\begin{eqnarray}
  \langle c_{\mathbf{k}\uparrow}c_{-\mathbf{k}+\mathbf{Q}'\downarrow} \rangle  \varpropto
  i\Delta_{\text{ddw}}\Delta_{\text{dsc}}f^2(\mathbf{k}).
  \end{eqnarray}
This indicates that the mixed state of DDW and DSC has a complementary
description as a mixed state of PDW and DSC. It is important to note
that the symmetry of the PDW order in this case is an extended
$s$-wave. It is also easy to see that the symmetries of the order in
the P-P channel and that  in the P-H channel  must be correlated
with each other: if one is the extended $s$-wave, the other is the
$d$-wave and vice versa. The above result holds for the DWCB order by
replacing $\mathbf{Q}'$ by $\mathbf{Q}$. In a global DSC state, the DWCB order must
exist when a checkerboard PDW order is present. The complementary connection suggests
that if the phase fluctuation leads to a Cooper pair modulation
pattern, orders in both the P-P and the
P-H channel have to be simultaneously considered in
microscopic models.

{\it BDG equation and co-existence of PDW and DWCB:}
 To verify the existence of the DWCB order and the above complementary connection,   we perform    a self-consistent BDG
 calculation.
We start from  the following general Hamiltonian on a two
dimensional square lattice  with the nearest-neighbor
attractive density interaction,
\begin{eqnarray}
\label{HH}
 H  = -\frac{1}{2}\sum_{\langle ij\rangle} \left[t_{ij}c^\dag_i c_j + h.c.\right] +
\sum_{\langle ij \rangle} V_{ij}n_{i} n_{j}  -\mu\sum_in_i.
\end{eqnarray}
The density interaction  $V_{ij}$   includes two parts,
\begin{equation}
V_{ij}= V^0_{}+\delta V_{\mathbf{r}_i,\mathbf{r}_j}
\end{equation}
where $V^0_{}$  favors a $d$-wave superconducting state and
$\delta V_{\mathbf{r}_i,\mathbf{r}_j}$ describes a modulating density interaction which
creates a small checkerboard modulation on the top of the uniform
superconducting state,
\begin{eqnarray}
 \delta V_{\mathbf{r},\mathbf{r}'}  = V'_{\mathbf{r}'-\mathbf{r}} ( \cos \mathbf{Q}\cdot \mathbf{r} + \cos
\mathbf{Q}\cdot \mathbf{r}'), \end{eqnarray} where  $V'_{\hat{x}} = V'$ and $V'_{\hat{y}}=-V'$
. The difference on the signs of  $V'_{\hat{x}}$ and $V'_{\hat{y}}$ provides us
the $d$-wave symmetry in the P-H channel order. We also note
that the sign difference  does not break rotational symmetry with
respect to a proper rotation center in the lattice.

Starting with Eq.~(\ref{HH}), we can derive the BdG equations by
introducing mean-field  decoupling of the nearest-neighbor interaction terms
and obtain a self consistent solution. 
The Bogoliubov-de Gennes equation is given by
\begin{eqnarray}
\left(
\begin{array}{cc}
\hat{H}_0 & \hat{\Delta}^* \\
\hat{\Delta} & -\hat{H}_0^*
\end{array}
\right)\left(
\begin{array}{c}
u_n(\mathbf{r})\\v_n(\mathbf{r})
\end{array}
\right)=E_n\left(
\begin{array}{c}
u_n(\mathbf{r})\\v_n(\mathbf{r})
\end{array}
\right)
\end{eqnarray}
where $\hat{H}_0$ and $\hat{\Delta}$ are transfer matrices such that
\begin{eqnarray}
\hat{H}_0 \psi_n(\mathbf{r}) &=&  -\left[{\sum}_{\delta} (t + W_{\delta}(\mathbf{r}))  + \mu (\mathbf{r})\right] \psi_n(\mathbf{r}+\delta)\\
\hat{\Delta} \psi_n(\mathbf{r}) &=& {\sum}_\delta \Delta_{\delta}(\mathbf{r})\, \psi_n(\mathbf{r}+\delta)
\end{eqnarray}
where $\psi_n(\mathbf{r})$ can be either $u_n(\mathbf{r})$ or $v_n(\mathbf{r})$, and $\delta$ denote nearest-neighbor vectors.  
The order
parameters are self-consistently determined by the self consistent
equations: 
the $d$-wave pairing amplitude on a bond $(\mathbf{r}, \mathbf{r}+\delta)$ is given by 
\begin{equation}
\Delta_{\delta}^{(1)}(\mathbf{r})= V^0 \langle c^{}_{\mathbf{r}\downarrow}
c_{\mathbf{r}+\delta\uparrow}^{} + c^{}_{\mathbf{r}+\delta\downarrow}
c_{\mathbf{r}\uparrow}^{}\rangle /2 ,
\end{equation}   
the pair density wave order in the
P-P channel is  
\begin{equation}
 \Delta_{\delta}^{(2)}(\mathbf{r}) = \delta
V_{\mathbf{r},\mathbf{r}+\delta} \langle c^{}_{\mathbf{r}\downarrow}
c_{\mathbf{r}+\delta\uparrow}^{} + c^{}_{\mathbf{r}+\delta\downarrow}
c_{\mathbf{r}\uparrow}^{}\rangle /2 ,
\end{equation} 
and  the density order in the
P-H channel is
\begin{equation}
 W_{\delta}(\mathbf{r}) = -\delta
V_{\mathbf{r},\mathbf{r}+\delta} \langle c_{\mathbf{r}\sigma}^\dagger
c_{\mathbf{r}+\delta\sigma}\rangle .
\end{equation}

 We have numerically solved  the BdG equation with various different parameter settings including different band structures in  the square lattice with different $N\times N$
sizes.
   We find that
the co-existence of the PDW and DWCD  and the symmetry
correspondence between
 them are the  robust results in our calculation for this system. For example,
with a parameter setting, $V^{0}_{} = 2.5t$ and $V' = t$, the
results are given by, for $\mathbf{r}'-\mathbf{r} = \hat{x}$ or $\hat{y}$,  $\Delta_{\mathbf{r}'-\mathbf{r}}^{(1)} (\mathbf{r}) = \pm \Delta_0$, $\Delta_{\mathbf{r}'-\mathbf{r}}^{(2)}(\mathbf{r}) =   \Delta_1 ( \cos \mathbf{Q}
\cdot \mathbf{r} + \cos \mathbf{Q}\cdot \mathbf{r}' )$,
and $W_{\mathbf{r}'-\mathbf{r}} (\mathbf{r})=  \pm W_0 (\cos \mathbf{Q} \cdot \mathbf{r} +\cos \mathbf{Q} \cdot \mathbf{r}')$ 
with $\Delta_0 =0.3t$, $\Delta_1 = 0.25 t$, and $W_0 = 0.15t$
, which are
corresponding to a state with
\begin{eqnarray}
    \Delta_{\text{dsc}}(\mathbf{r}) &=& \Delta_0 \\
    \Delta_{\text{pdw}} (\mathbf{r}) &=& \Delta_1 \cos \mathbf{Q} \cdot \mathbf{r} \\
    W_{\text{dwcb} } (\mathbf{r}) &=& W_0 \cos \mathbf{Q} \cdot \mathbf{r}
\end{eqnarray}
We note that PDW and DWCB has the same spatial modulation
 as $\cos \mathbf{Q}\cdot \mathbf{r}$, while the symmetries of the PDW and DWCB orders are the extended $s$-wave and $d$-wave
 respectively. As shown in Fig.~\ref{fig_1},
 DWCB  has $4a\times 4a$ periodicity and $d$-wave symmetry.
Similar order parameters have been mentioned in Ref.\cite{PODOLSKY2003}.
 The   solutions   are independent of the
initial guesses for the local variables and converge quickly as $N$
increases.

  After demonstrating the co-existence of the DWCB and PDW, now we are interested in the physical effects of the DWCB order.
   To obtain a clear picture of the DWCB, we 
  illustrated a static pattern of the bond strength of the DWCB order in Fig.~\ref{fig_1}:
  \begin{eqnarray}
  W_{\hat{x}} (\mathbf{r}) &=&  \text{Re} (W_0) \left[ \cos \frac{\pi x}{2} - \sin \frac{\pi x}{2} + \cos \frac{\pi y}{2} \right] \\
  W_{\hat{y}} (\mathbf{r}) &=&  - \text{Re} (W_0) \left[ \cos \frac{\pi y}{2} - \sin \frac{\pi y}{2} + \cos \frac{\pi x}{2} \right]
  \end{eqnarray}
  It is clear that the DWCB order has $4a\times 4a$ periodicity and $d_{x^2-y^2}$ symmetry.
Similar order parameters have been mentioned in Ref.\cite{PODOLSKY2003}.

\begin{figure}
      \includegraphics[width=9cm]{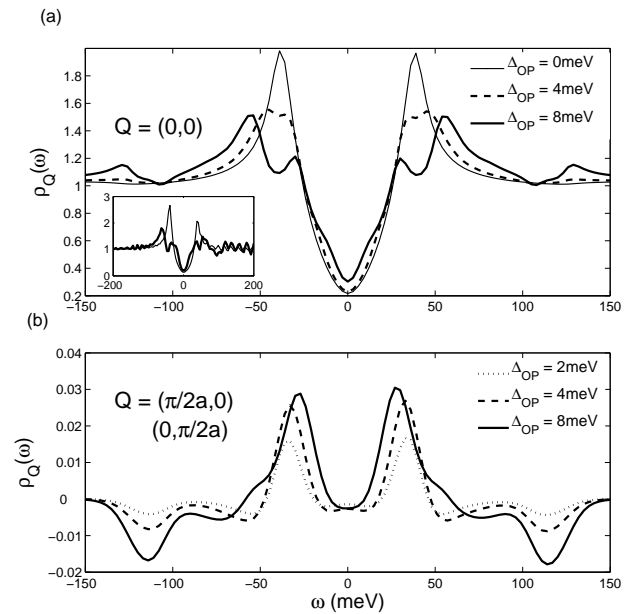}
\caption{\label{fig_2}(a) The average DOS in the particle-hole symmetric case. The inset shows the average DOS with the finite chemical
potential included in the band dispersion provided by Norman {\it et
al}.~\cite{NORMAN1995}. $\Delta_{\text{OP}}$ represents $W_0$. (b) The Fourier components of LDOS at $\mathbf{Q} = \{(
\pi/2,0),(0,\pi/2)\}$.}
\end{figure}

Now we numerically calculate the average density of
states(DOS), $\rho(\omega)$, and the Fourier components, $\rho_\mathbf{Q}
(\omega)$, at the wavevectors of the DWCB order: $\mathbf{Q} =
\{(\pi/2,0),(0,\pi/2)\}$, and directly compare them with experimental
results. We calculate the above quantities in two situations with
different band dispersions. The results are rather general and are
insensitive to the bare band structures. First, we performed
calculations in the particle-hole symmetric case. For a simple band
dispersion, we choose $t = -125$meV and $\mu = 0$. $\Delta_{0}
= 40$meV, which is relevant for underdoped BSCCO. The imaginary part
of the self energy $\eta = 5$meV is used for the entire numerical
calculation. Fig.~\ref{fig_2}(a) shows the average DOS($\mathbf{Q}=(0,0)$)
normalized by the non-interacting Fermi liquids. In the absence of
DWCB, there are  sharp coherence peaks at the
energy of superconducting gap. As DWCB order develops, the coherence
peaks are suppressed and pushed away while the spectrum at low
energy remains unchanged. Fig.~\ref{fig_2}(b) shows the Fourier
components of the local density of states(LDOS) at the wavevectors, $\mathbf{Q}$.
As expected, $\rho_\mathbf{Q} (\omega)$ is even with respect to $\omega$,
namely, $  \rho_\mathbf{Q} (\omega)=\rho_\mathbf{Q} (-\omega).$ Second, we repeat our
calculations with the band dispersion provided by Norman {\it et
al.}~\cite{NORMAN1995} and the result is displayed at the inset in
Fig.~\ref{fig_2}(a). The band energy dispersion is now modified as such
\begin{eqnarray}
 \xi_\mathbf{k} &=& t_1/2 (\cos k_x + \cos k_y ) + t_2 \cos k_x \cos k_y \nonumber\\
      &+& t_3/2 (\cos 2 k_x + \cos 2 k_y ) \nonumber\\ 
      &+& t_4/2 ( \cos 2 k_x \cos k_y + \cos k_x \cos 2 k_y ) \nonumber\\
      &+& t_5 \cos 2k_x \cos 2 k_y -\mu ,
\end{eqnarray}
where $t_1 = -0.5951$eV, $t_2 = 0.1636$eV, $t_3 = -0.0519$eV,
$t_4 = -01117$eV, and $t_5 = 0.0510$eV. The chemical potential $\mu$
is now set to -0.1660eV. Compared with the particle-hole symmetric
case, the effect of DWCB on the LDOS is insensitive to the energy
band structure. Qualitatively the numerical results are strikingly
consistent with experimental results~\cite{HOWALD2003,FANG2006}, and
the large gap region can be represented by the presence of DWCB
which is weak at 8-12meV.

 Analytically, the general features in STM measurements can
be captured by the DWCB. First, due to the anisotropy inherited from
the $d$-wave factor of pairing, a weak DWCB order has a much
stronger effect on the antinodal region than on the nodal region,
thus naturally explaining the puzzling dichotomy between the nodal
and antinodal excitations in high-temperature superconductors: The
local phase fluctuations of Cooper pairs lead to a local modulation
of $d$-wave ordering in the particle-hole channel, which strongly
affects the antinodal single particle excitations. Second, like PDW,
DWCB is bond-centered and consequently, $\rho_\mathbf{Q} (\omega)$ is an even
function of $\omega$, too\cite{CHEN2004}. The symmetry has been shown
to distinguish the PDW from the typical particle-hole CDW. The
existing experimental results are consistent with the even case.

The above result demonstrates the consistency between the DWCB order
and the STM experimental results in superconducting state. Now we
show that the DWCB order also captures important physics in the
pseudogap phase. One important feature of the pseudogap phase is the
non-dispersive Fermi arc developed from the nodal point along Fermi
surface observed in ARPES~\cite{VERSHININ2004}. The Fermi arc has
been used to explain the STM result in the pseudogap
phase\cite{Chatterjee2006}. If the pseudogap phase is strongly
connected to phase fluctuations of $d$ wave superconductivity, the
single particle spectrum should reflect the DWCB order. Therefore, a
robust Fermi arc feature should exist in the mixed DWCB and DSC
phases in high temperature. We found that this is indeed the case.

\begin{figure}
  \includegraphics[width=8.5cm]{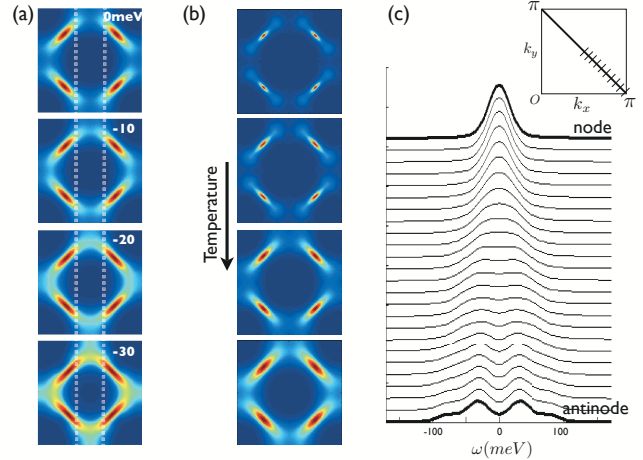}
  \caption{\label{fig_3}(a) With DSC and DWCB coexisted, the spectral weights, $A (\mathbf{k}, \omega)$, are plotted as a function of $\omega$ (-30meV $\sim$ 0meV) in the first Brillouin zone.
  Here we used $t=-125$meV, $t'=\mu=0$meV, and $T = 120K$. Each BZ is segmented
  by 100$\times$100. White dotted lines indicate the non-dispersive Fermi arcs. 
  (b) Temperature dependence of $A(\mathbf{k},\omega)$ at the Fermi level, $\omega = 0$meV.
  (c) EDC curves from the nodal to antinodal point along the Fermi surface show the gapless region, or Fermi arc. }
\end{figure}

In Fig.~\ref{fig_3}, we have calculated $A(\mathbf{k},\omega)$ in the pseudogap
state within the model of Franz and Millis~\cite{FRANZ1998}. Fig.~\ref{fig_3}(a) shows the numerical solutions of the
spectral weight, $A(\mathbf{k},\omega)$, as a function of $\omega$ at high
temperature $T=120$K where $\Delta_{0} = 40$meV and the DWCB order
is equal to 8meV. As expected in the pseudogap state, the scattering
vector connecting the tips of Fermi arcs unchanges as energy
$\omega$ increases, and is nearly equal to wavevectors of the
$d$-checkerboard order parameter, $|\mathbf{Q}| = \pi/2$. Fig.~\ref{fig_3}(c) shows the spectral weight at $\omega=0$ for the  cuts
perpendicular to the Fermi arc which matches the experimental
results\cite{Chatterjee2006}. For the temperature dependence, the
length of the Fermi arc is linearly increasing as the temperature
rises above Tc~\cite{KANIGEL2006}. As seen in Fig.~\ref{fig_3}(c), at very low temperature the Fermi
surface is gapped except for nodal point, $( \pi/2, \pi/2)$. As the
temperature rises, the gapless region elongates along the Fermi
surface, with slight broadening in the direction perpendicular to
the Fermi surface.

In summary, the DWCB order offers a unified explanation for the STM experiments in both
the superconducting and pseudogap phases. A number of important issues need to be addressed.
 First, while we show that the complementary connection exists between
the $d$-wave order in the particle-hole channel and the pair density wave order in the
particle-particle channel, the order wavevectors should be determined by microscopic models.
 In general, the pair fluctuation in different
mircoscopic models can lead to different
order wavevectors in the particle-hole channel. For
example, the pair fluctuation can be anisotropic in the space that breaks
the rotational lattice symmetry, and will result in a stripe-like
one dimensional ordering. Second, it is interesting that in
a recent renormalization group study of the electron
phonon interaction in cuprates, the authors has shown that the DWCB
order rises from the coupling to the half breathing or the
$B_{1g}$ phonons~\cite{DHLEE2005}. Combining their results with ours suggests
that the superconducting phase fluctuations may be strongly coupled with phonons.
This hypothesis is of great importance and in need of future investigation.
Finally, although a full calculation based on the BDG equation with the DWCB order is yet to be completed,
preliminary results show that the qualitative conclusion drawn here should remain valid.

 In conclusion,  the order in particle hole channel, i.e., the DWCB
order,  can explain the STM spectrum and the
nodal-antinodal dichotomy observed in both STM and ARPES
experiments~\cite{ZHOU2004,MCELROY2005A}. The presence of the DWCB order also preserves the gapless dispersion at nodal points and simultaneously generates a Fermi arc with little dispersion around nodal
points at finite temperature. The results are consistent with ARPES experiments and provide an explicit physical explanation of the fermi arc in the pseudogap phase.

J.P.Hu would like to thank Wei-Sheng Lee for helpful discussions. J.
P.Hu and K.Seo are supported by Purdue research funding. H.D.Chen is
supported by the U.S. Department of Energy, Division of Materials
Sciences under Award No. DEFG02-91ER45439, through the Frederick
Seitz Materials Research Laboratory at the University of Illinois at
Urbana-Champaign.

\bibliography{dwcb}

\end{document}